\documentclass[12pt]{article}

\usepackage[english]{babel}

\usepackage[letterpaper,top=2cm,bottom=2cm,left=3cm,right=3cm,marginparwidth=1.75cm]{geometry}

\usepackage{amsmath}
\usepackage{graphicx}
\usepackage[colorlinks=true, allcolors=blue]{hyperref}
\usepackage{amssymb}
\usepackage[utf8]{inputenc}
\usepackage{lipsum}
\usepackage{multicol}
\usepackage{physics}

\title{Gravity from Invariant Weyl-Integrable space-time (IWIST) }
\author{Jos\'e Edgar Madriz Aguilar$^{1}$$^{\dagger}$,  A. Bernal $^{2}$, M. Montes $^{1}$, J. Zamarripa $^{1}$ }

\begin{document}
	\date{}
	\maketitle
\begin{center}
		$^{1}$ Universidad de Guadalajara (UdG), Centro Universitario de Ciencias Exactas e ingenier\'{i}as,
        Departamento de Matem\'{a}ticas, 
		 Av. Revoluci\'on 1500 S.R. 44430, Guadalajara, Jalisco, M\'exico,\\
		and\\
		$^{2}$ Universidad de Guadalajara, Centro Universitario de los Valles, Departamento de Ciencias Naturales y Exactas,\\
        Carretera Guadalajara-Ameca Km 45.5\\
		C.P. 46600, Ameca, Jalisco, México \\
		\vspace{0.3cm}
		$^{\dagger}$ Corresponding author: José Edgar Madriz Aguilar\\
		E-mails:  jose.madriz@academicos.udg.mx, 
		alfonso.bernal@academicos.udg.mx,
		mariana.montes@academicos.udg.mx,
        jose.zamarripa@academicos.udg.mx
\end{center}
	\begin{abstract}
		In this paper we propose a geometrically motivated formulation of gravity in an invariant Weyl--Integrable space-time (IWIST), in which the background geometry is not imposed a priori but is dynamically determined through the Palatini variational principle. For a scalar field non-minimally coupled to gravity action, the resulting compatibility condition defines a Weyl--Integrable geometry which is preserved by a local Weyl transformation of the metric and the geo\-me\-tri\-cal scalar field. We construct an action invariant under both diffeomorphisms and Weyl transformations and introduce a Weyl-covariant variational procedure based on an invariant extension of the divergence theorem. The corresponding field equations are obtained for the metric, the Weyl scalar, and the Weyl gauge field. We further formulate the theory in the Einstein--Riemann frame, where the effective metric is Riemannian and the scalar field is re\-inter\-pre\-ted as a phy\-si\-cal degree of freedom of geometric origin. Finally, we propose a geometrical mechanism for the explicit breaking of Weyl symmetry by introducing a coupling between the Weyl gauge field and a current constructed from the scalar sector. The resulting theory remains diffeomorphism covariant while departing from local Weyl invariance. The Einstein--Riemann representation of the broken theory contains an additional generally covariant interaction, making the Weyl-invariant model a particular limit of a more general gravitational theory.
        
	\end{abstract}

	\\
	\vspace{0.3cm}
	
	
	\section{\large{Introduction}}

Since Weyl introduced his unified theory of gravitation and electromagnetism based on a non-Riemannian geometry in 1918 \cite{Weyl1918}, Weyl geometry has attracted continuous attention as a possible geometrical framework for gravitation. In Weyl's original formulation, the non-metricity condition is determined by an arbitrary one-form field, leading to the well-known second-clock effect pointed out by Einstein shortly after the theory was proposed. Although this criticism prevented the original Weyl theory from becoming a viable theory of gravitation, it also motivated the search for restricted versions of Weyl geometry in which this undesirable feature is absent.\\

A particularly important realization is the Weyl Integrable Space-Time (WIST), where the Weyl one-form is assumed to be an exact differential, $\sigma_{\mu}=\partial_{\mu}\phi$. In this case, the second-clock effect disappears and the affine structure remains compatible with a generalized notion of length transport. Early cosmological applications of WIST were proposed by Novello and collaborators, who explored the possibility that integrable Weyl geometry could provide a natural framework for non-singular cosmological models and geometrically induced phase transitions between different space-time configurations \cite{Novello1983,Novello1992,Oliveira1997}. These works established that the scalar field associated with the Weyl geometry could acquire a genuine geometrical interpretation rather than representing an independent matter degree of freedom.\\

During the last two decades, the Brazilian school led by Romero and collaborators has considerably extended the mathematical and physical foundations of WIST. One of their most important achievements was the demonstration that General Relativity can be reformulated entirely within Weyl integrable geometry without changing its physical predictions \cite{Romero2012}. In this formulation, the gravitational theory becomes explicitly invariant under Weyl transformations and introduces the concept of \emph{Weyl frames}, showing that different geometrical representations correspond to the same physical phenomena. This viewpoint was further developed in the analysis of conformally flat space-times, where the gravitational interaction may alternatively be interpreted as arising from the Weyl scalar field defined on a flat Minkowski background \cite{RomeroCF2012}. Such results considerably strengthened the geometrical interpretation of scalar degrees of freedom in gravitation and clarified the relation between Weyl geometry and scalar--tensor theories.\\

The invariant formulation of gravity in WIST also motivated several extensions in different directions. The relation between WIST and Brans--Dicke theory was investigated from a geometrical viewpoint, emphasizing that the scalar field appearing in the Einstein frame may be regarded as part of the affine structure rather than as an independent physical field \cite{Romero2012}. The concept of proper time was subsequently revisited within the Ehlers--Pirani--Schild axiomatic framework, leading to a consistent definition of clock measurements in Weyl integrable geometry \cite{Avalos2018}. Likewise, the Hawking--Penrose singularity theorems were generalized to scalar--tensor theories formulated on WIST manifolds, showing that many global properties of General Relativity admit natural extensions to Weyl geometry \cite{Lobo2015}. Lower-dimensional gravitational models were also investigated, revealing that WIST possesses a well-defined Newtonian limit in $(2+1)$ dimensions, together with vacuum solutions that have no direct analogue in Einstein gravity \cite{Madriz2015}.\\

Beyond these formal developments, WIST has also been employed in several cosmological scenarios. Different authors have investigated isotropic and anisotropic cosmologies, accelerated expansion driven by the Weyl scalar field, interacting dark energy models, and inhomogeneous cosmological solutions, illustrating the flexibility of the geo\-me\-tri\-cal framework in describing different stages of the cosmic evolution \cite{Miritzis2004,Paliathanasis2020,Chatzidakis2022}. These investigations reinforce the idea that the scalar field emerging from Weyl integrable geometry may provide an alternative geometrical description of phenomena that are usually attributed to additional matter sources.\\

Despite the Weyl integrable geometry has been successfully employed to formulate a wide variety of gravitational and cosmological models, an important conceptual issue deserves further attention. Most formulations begin by constructing an action from the Weyl scalar curvature supplemented by an explicit kinetic term for the geometrical scalar field \cite{Novello1983,Oliveira1997,Romero2012,Miritzis2004,Paliathanasis2020,Chatzidakis2022}. While the Weyl scalar curvature possesses the appro\-pria\-te conformal weight to guarantee the invariance of the purely geometrical contribution to the action, the same property is not automatically shared by an independently introduced kinetic term proportional to $g^{\mu\nu}\phi_{,\mu}\phi_{,\nu}$. Since $\phi$ is an ordinary scalar field, its covariant derivative coincides with its partial derivative independently of the affine connection, namely $\nabla_{\mu}\phi=\partial_{\mu}\phi$.  Consequently, under the standard Weyl transformations,
\[
\bar g_{\mu\nu}=e^{f(x)}g_{\mu\nu},
\qquad
\bar\phi=\phi+f(x),
\]
an independent kinetic contribution generally generates additional terms proportional to $\phi_{,\mu}f^{,\mu}$ and $f_{,\mu}f^{,\mu}$, which do not cancel identically nor reduce, in general, to a total divergence. This observation suggests that the Weyl invariance of the complete action cannot be regarded as a direct consequence of the geometrical invariance of the compatibility condition alone, but rather requires additional assumptions concerning the construction of the action and, possibly, the underlying variational principle. These considerations motivate the present work, where the geometrical symmetry associated with the compatibility condition is taken as the fundamental guiding principle for cons\-truc\-ting both the action and the corresponding field equations.\\

 On the other hand, in many formulations of scalar-tensor theories of gravity, the action is assumed to be invariant only under diffeomorphisms. This is because the background geometry is considered Riemannian.  Nevertheless, the dy\-na\-mi\-cal description is usually developed in two distinct conformal frames. In the Jordan frame, the scalar field is non-minimally coupled to the gravitational sector, whereas in the Einstein frame it is minimally coupled. The latter is obtained from the former by means of a conformal transformation of the metric. In the Einstein frame, the gravitational field equations acquire a form analogous to those of General Relativity, while the scalar field is interpreted as an effective matter degree of freedom. The coexistence of these two frames leads to the so-called \emph{frame problem}. This issue stems from the lack of a ge\-ne\-ra\-lly accepted criterion for determining which frame should be regarded as the physical one. Consequently, it remains unclear in which frame the field equations provide the most appropriate des\-crip\-tion of gravitational phenomena \cite{IQMA}. Some interesting papers on scalar-tensor theories of gravity in which conformal transformations relating the Jordan and Einstein frames are a standard tool in scalar--tensor and Brans--Dicke theories are \cite{U1,U2,U3,U4,U5,U6,U7,U8}. \\

Something that is often overlooked in scalar-tensor theories of gravity is the fact that a conformal transformation acting solely on the metric, generally modifies the under\-lying geometric structure \cite{U8A,U8B}. Consequently, the background geometry is no longer the same after the transformation is applied. To better understand this issue, it is convenient to recall first how a spacetime  geometric structure is characterized.  A spacetime geo\-me\-try is characterized by the metric tensor and the affine connection, or equivalently, by the corresponding nonmetricity and torsion tensors. The nonmetricity characterizes the compatibility between the metric and the affine connection, whereas the torsion encodes the symmetry properties of the affine connection. Therefore, any transformation that modifies these geometric objects changes the underlying geometric structure. In particular a conformal transformation of the metric, in a geometry alters the non-metricity or metricity, depending on the case, and as a consequence the underlying geometry changes. Thereby, it becomes clear that a conformal transformation acting solely on the metric modifies the geometric structure of spacetime. Thus, if the Jordan frame is formulated on a Riemannian manifold, the geometry naturally associated with the Eins\-tein frame should, in general, be regarded as non-Riemannian, since the compatibility relation between the metric and the affine connection is altered by the transformation. \\

As a consequence of the change in the background geometry induced by a conformal transformation of the metric, observers that follow geodesics in the original Jordan frame become, in general, non-geodesic in the geometry associated with the Eins\-tein frame. As a result, the description of gravitational phenomena in the Eins\-tein frame involves effective force terms that are absent in the original frame that are originated exclusively from the geometric transformation. Therefore, the physical measurements performed by such observers are affected by these additional contributions. This fact strongly suggests that, once the change in the underlying geometry is pro\-per\-ly accounted for, the Jordan and Einstein frames should not be considered physically equivalent.\\

An intriguing feature of a broad class of scalar--tensor theories of gravity is their close connection with Weyl-integrable geometry. When the Palatini variational principle is applied to actions containing a non-minimal coupling between a scalar field and the curvature, the affine connection is determined dynamically rather than being prescribed \emph{a priori}, and the resulting background geo\-me\-try is naturally of the Weyl-integrable type instead of the Riemannian one. This result suggests that Weyl-integrable geometry provides the natural geometrical framework for scalar--tensor theories and, more ge\-ne\-ra\-lly, for gravitational theories involving scalar fields non-minimally coupled to gravity \cite{U9,U10,U11,U12}.\\

In this paper, we propose a geometrically consistent formulation of an Einstein--Hilbert action containing a scalar field non-minimally coupled to the curvature based on a simple guiding principle: the symmetries of the gravitational action should coincide with the symmetries of the dynamically generated background geometry. Starting from gravitational theories with a non-minimal coupling between the scalar field and the curvature, the background geometry is obtained through the Palatini variational principle rather than being imposed \emph{a priori}. The compatibility condition resulting from the variational procedure then determines the corresponding geometrical symmetry group, under which the gravitational action is required to remain invariant. Consequently, the action, the variational principle and the field equations become manifestations of the same underlying geometrical structure,
providing a natural framework for addressing the frame problem. \\

More explicitly, the central idea is that the action of a gravitational theory as described above, should determine its own underlying geometry through the Palatini variational principle, rather than assuming a Riemannian geometry \emph{a priori}. Furthermore, we identify the group of geometrical transformations that leaves the compatibility condition between the metric and the affine connection invariant, and use this symmetry group to construct a scalar action that is itself invariant under these transformations. In other words, if a scalar quantity in a non-Riemannian geometry is defined as one that remains invariant under the group of transformations preserving the compatibility condition, then in order to have a gravitational action being a scalar, it must also be invariant under the same group of geometrical symmetries characterizing the background geometry. Finally, we introduce suitable modifications to the variational procedure in order to derive field equations that remain invariant under the same group of geometrical transformations, with particular emphasis on the case of a scalar field non-minimally coupled to gravity.\\

To our purposes, the paper is organized as follows. In section 1 we give an introduction. In section 2 we propose an invariant action under the Weyl group of symmetries. In section 3 we present a modified version of the divergence theorem, which is invariant under the Weyl symmetries. Section 4 is devoted to the derivation of the field equations associated to the Weyl invariant action constructed in section 2, by means of the implementation of the new version of the divergence theorem. In section 5 we introduced the Einstein-Riemann frame and obtain the field equations in this Riemannian framework. In section 6 we propose a mechanism for the explicit breaking  of the Weyl symmetry, obtaining as a result a gravitational theory that departs from local Weyl invariance. Finally in section 7 we give some concluding remarks.

	\section{The Invariant action in the Weyl frame}
  Let us start with an action, very common in general relativity, that describes a scalar field non-minimally coupled to gravity in the form
\begin{equation}\label{eq1}
    S=\int d^4x\,\sqrt{-g} \left[\frac{1}{2\kappa}\,\Omega(\Phi)R-\frac{1}{2}g^{\mu\nu}\partial_{\mu}\Phi\partial_{\nu}\Phi+V(\Phi)\right],
\end{equation}
where $\kappa=(8\pi G)$ is the usual gravitational coupling constant that appears in the Einstein-Hilbert action, $\Phi(x^{\lambda} )$ is a scalar field defined on the space-time manifold , $\Omega(\Phi)$ is a differentiable well-behaved function of $\Phi$ that determines the non-minimal coupling of $\Phi$ with gravity, and $V(\Phi)$ is the scalar potential associated to $\Phi$. \\

If we consider  that the metric and the affine connection in \eqref{eq1} are independent quantities, which means that we are not considering apriori an initial background geometry for the action \eqref{eq1}, and we adopt the Palatini's variational principle, the variational of the action \eqref{eq1} with respect to the affine connection leaves to the compatibility condition  
\begin{equation}\label{eq2}
    \nabla_{\mu}g_{\alpha\beta}=\partial_{\mu}[-\ln \Omega(\Phi)]\,g_{\alpha\beta}.
\end{equation}
By means of the transformation $\varphi=-\ln (\Omega(\Phi))$, the condition \eqref{eq2} reads 
\begin{equation}\label{c3}
     \nabla_{\mu}g_{\alpha\beta}=\partial_{\mu}\varphi\,g_{\alpha\beta}.
\end{equation}
The compatibility condition \eqref{c3} is invariant under the set of transformations
\begin{eqnarray}
    \bar{g}_{\alpha\beta}=e^{f(x)}g_{\alpha\beta}\,,\label{wt1}\\
    \bar{\varphi}=\varphi + f(x)\,.\label{wt2}
\end{eqnarray}
This set of transformations is known as the Weyl transformation group or Weyl group of symmetries. Now, one natural step is rewrite the action \eqref{eq1} in terms of  the redefined field $\varphi$. Hence,  it follows from $\Omega (\Phi)=e^{-\varphi}$ that the field $\varphi$ can be expressed in the form
\begin{equation}\label{c4}
    \Phi=\Omega^{-1}\left[e^{-\varphi}\right]\,,
\end{equation}
where $\Omega^{-1}$ represents the inverse function of $\Omega$, that satisfies $\Omega^{-1}(\Omega (x))=x$.  Thus, employing the chain rule it follows from \eqref{c4} that
\begin{equation}\label{c5}
    \partial_{\mu}\Phi=\partial_{\varphi}\left[\Omega^{-1}\left(e^{-\varphi}\right)\right]\,\partial_{\mu}\varphi\,.
\end{equation}
With the help of \eqref{c5} the kinetic term in the action \eqref{eq1} becomes
\begin{equation}\label{c6}
    \frac{1}{2}g^{\mu\nu}\partial_{\mu}\Phi\partial_{\nu}\Phi = \frac{1}{2} g^{\mu\nu}\left(\partial_{\varphi}\left[\Omega^{-1}\left(e^{-\varphi}\right)\right]\right)^{2}\partial_{\mu}\varphi\,\partial_{\nu}\varphi= \frac{1}{2}\Gamma(\varphi)g^{\mu\nu}\partial_{\mu}\varphi\partial_{\nu}\varphi \,,
\end{equation}
where
\begin{equation}\label{c7}
    \Gamma(\varphi)= \left(\partial_{\varphi}\left[\Omega^{-1}\left(e^{-\varphi}\right)\right]\right)^{2}\,.
\end{equation}
At this point it is important to note that as $\varphi$ transforms according to the rule $\bar{\varphi}=\varphi+f$, then results valid the transformation rule
\begin{equation}\label{c8}
    \bar{\Omega}(\bar{\Phi})=e^{-f}\,\Omega(\Phi).
\end{equation}
The expression \eqref{c8} indicates that $\Gamma(\varphi)$ does not transform under the Weyl group as an scalar. For example, if we consider $\Omega(\Phi)=\vartheta \phi^n$, with $\vartheta$ being a coupling constant, and $n>0$, it is not difficult to show that in this particular case $\Omega^{-1}(e^{-\varphi})=\vartheta^{1/n}e^{-\varphi/n}$ and thus $\Gamma(\varphi)=-(\vartheta^{1/n}/n)e^{-\varphi/n}$. Hence, with the help of \eqref{wt2} or equivalently \eqref{c8},  we arrive to the transformation rule $\bar{\Gamma}(\bar{\varphi})=e^{-f/n}\Gamma(\varphi)$, which explicitly shows that $\Gamma(\varphi)$ is not weyl-invariant. Of course it is important to keep in mind that in this development we have considered $\bar{\Gamma}(\bar{\varphi})=\Gamma (\bar{\varphi})$.   \\

Continuing with our calculation, with the help of 
\eqref{c4}, \eqref{c6} and \eqref{c7}, the action \eqref{eq1} finally written in terms of $\varphi$ results to be
\begin{equation}\label{c9}
    S=\int d^4x\,\sqrt{-g} \left[\frac{1}{2\kappa} e^{-\varphi}\,R-\frac{1}{2}\Gamma(\varphi)g^{\mu\nu}\partial_{\mu}\varphi\partial_{\nu}\varphi+\tilde{V}(\varphi)\right], 
\end{equation}
where $ \tilde{V}(\varphi)=V[\Phi(\varphi)]$. It is important to mention here that a similar conclusion to the one obtained for $\Gamma(\varphi)$ results valid in the case of the scalar potential. If we consider for example $\tilde{V}(\varphi)=(1/2)m^2\varphi^2$, then $\bar{\tilde{V}}(\bar{\varphi})=\tilde{V}(\bar{\varphi})=(1/2)m^2(\varphi +f)^2\neq \tilde{V}(\varphi)$, and thus the potential is not weyl-invariant as well.\\

In addition, it is not difficult to see that  the kinetic term in \eqref{c9} is not invariant under the Weyl group of transformations \eqref{wt1} and \eqref{wt2}. Moreover, taking into account the transformation rules $\sqrt{-\bar{g}}=e^{2f}\sqrt{-g}$ and $\bar{R}=e^{-f}R$,   neither is the complete action \eqref{c9} . It means that the action \eqref{c9} is not a scalar i.e. it is not an invariant under the group of symmetries of the background geometry, which as we mentioned before, is of the Weyl-Integrable type, and from this arises the need to construct a truly scalar action. \\

In this manner, we propose the new Weyl-scalar action 
\begin{equation}\label{c10}
    S=\int d^4x\,\sqrt{-g}\,e^{-\varphi}\left[\frac{1}{2\kappa} R-\frac{1}{2}\omega(\varphi)\,g^{\mu\nu}{\cal D}_{\mu}\varphi{\cal D}_{\nu}\varphi+e^{-\varphi}U(\varphi)-\frac{1}{4}e^{\varphi}\Psi_{\mu\nu}\Psi^{\mu\nu}\right]
\end{equation}
where $\omega(\varphi)$ and $U(\varphi)$ are Stueckelberg field functions, that must obey the transformation rules 
\begin{eqnarray}\label{c11}
    \bar{\omega}(\bar{\varphi})&=&\omega(\bar{\varphi}-f)=\omega(\varphi),\\
    \label{c12}
    \bar{U}(\bar{\varphi})&=& U(\bar{\varphi}-f)=U(\varphi)
\end{eqnarray}
An important feature in the formulation of \eqref{c10} is the introduction of the gauge derivative
\begin{equation}\label{c12a}
    {\cal D}_{\mu}=\partial_{\mu}+\gamma B_{\mu},
\end{equation}
being $B_{\mu}$ the coordinate components of a one-form field, here referred to as the \emph{Weyl gauge field}, defined at each point of space-time, and $\gamma$ a dimensional constant. In order for the action \eqref{c10} to be invariant under the Weyl transformation group \eqref{wt1}-\eqref{wt2}, the following condition is required
\begin{eqnarray}
    {\cal D}_{\mu}\bar{\varphi} &=& {\cal D}_{\mu}\varphi,\nonumber\\
    \partial_{\mu}\bar{\varphi}+\gamma\bar{B}_{\mu}\bar{\varphi} &=& \partial_{\mu}\varphi+\gamma B_{\mu}\varphi,\nonumber\\
    \gamma\bar{\varphi}\bar{B}_{\mu}&=& \gamma\varphi B_{\mu}-\partial_{\mu}f,\nonumber\\
    \bar{\varphi}\bar{B}_{\mu} &=& \varphi B_{\mu}-\gamma^{-1}\partial_{\mu}f. \label{c13}
\end{eqnarray}
In the action \eqref{c10} the  tensor $\Psi_{\mu\nu}$ is named the gauge strength tensor and it was introduced to give dynamics to the gauge field $B_{\mu}$. The last term in \eqref{c10} represents the kinetic energy of the gauge field. The gauge strength tensor is defined as
\begin{equation}\label{c14}
    \Psi_{\alpha\beta}=\partial_{\alpha}(\varphi B_{\beta})-\partial_{\beta}(\varphi B_{\alpha}).
\end{equation}
With the help of \eqref{c13}, the gauge strength tensor under the Weyl group obeys the transformation rule
\begin{eqnarray}
    \bar{\Psi}_{\alpha\beta} &=& \partial_{\alpha}(\bar{\varphi}\bar{B}_{\beta})-\partial_{\beta}(\bar{\varphi}\bar{B}_{\alpha}),\nonumber\\
    &=& \partial_{\alpha}(\varphi B_{\beta})-\gamma^{-1}\partial_{\alpha}\partial_{\beta} f-\partial_{\beta}(\varphi B_{\alpha})+\gamma^{-1}\partial_{\beta}\partial_{\alpha}f,\nonumber\\
    &=& \partial_{\alpha}(\varphi B_{\beta})-\partial_{\beta}(\varphi B_{\alpha}),\nonumber\\
    \bar{\Psi}_{\alpha\beta} &=& \Psi_{\alpha\beta} ,\label{c15}
\end{eqnarray}
which indicates that $\Psi_{\mu\nu}$ is invariant under the Weyl group. It is important to clarify that in the exppresions \eqref{c11} and \eqref{c12} we have defined $\omega(\varphi)=\Gamma(\varphi)$ and $U(\varphi)=\tilde{V}(\varphi)$. \\

We would like to emphasize that, compared with the action \eqref{c9}, the action \eqref{c10} contains additional considerations.  In order to the transformations \eqref{c11} and \eqref{c12} to be valid, necessarily $\bar{\Gamma}(\bar{\varphi})\neq \Gamma(\bar{\varphi})$ and $\bar{\tilde{V}}(\varphi)\neq \tilde{V}(\bar{\varphi})$. However, we must remember that in the action \eqref{c9} it was assumed that $\bar{\Gamma}(\bar{\varphi})=\Gamma(\bar{\varphi}) $ and $\bar{\tilde{V}}(\bar{\varphi})=\tilde{V}(\bar{\varphi})$, which are different considerations. This is due to the fact that in \eqref{c9} the expressions $\Gamma(\varphi)$ and $\tilde{V}(\varphi)$ are only functions while in \eqref{c10} the quantities  $\omega(\varphi)=\Gamma(\varphi)$ and $U(\varphi)=\tilde{V}(\varphi)$ are Stueckelberg field functions. 

\section{ A modified divergence theorem}

As is well known, in the derivation of the gravitational field equations from the action of a gravitational theory, boundary conditions play a fundamental role. During the variational procedure, boundary terms naturally arise and are handled through the integral form of Gauss's divergence theorem, which relates the volume integral of the divergence of a vector field to the corresponding flux through the boundary hypersurface. Consequently, the integral form of this theorem is of crucial importance in determining the final form of the field equations. However, since our aim is to formulate a theory in which not only the action but also the resulting field equations are invariant under the Weyl group of transformations, it is necessary to require that the integral form of the divergence theorem itself also possesses Weyl invariance.\\

In order to do so, we begin by recalling the Gauss divergence theorem in an $n$-dimensional Lorentzian manifold which states the following:\\

{\bf Theorem}: {\it Let $(M,g)$ be an $n$-dimensional Lorentzian manifold endowed with a metric $g$ and an affine connection $\nabla$ satisfying the metric compatibility condition
\begin{equation}\label{eq2a}
    \frac{Dg}{d\sigma}(\mathbf{u},\mathbf{v})=0,
\end{equation}
where $\lambda(\sigma):I\subseteq\mathbb{R}\rightarrow M$ is a smooth curve, and $\mathbf{u},\mathbf{v}\in T_{\lambda(\sigma)}M$. If $\mathbf{A}=A^{\mu}\partial_{\mu}\in\chi(M)$ is a smooth vector field, then, in any coordinate chart,
\begin{equation}\label{eq3}
    \int_{V} d^{n}x\,\sqrt{-g}\,\nabla_{\mu}A^{\mu}
    =
    \oint_{\partial V}
    A^{\mu}n_{\mu}\,
    \sqrt{|\gamma|}\,
    d^{(n-1)}x,
\end{equation}
where $\gamma=\det(\gamma_{ij})$, with $i,j=1,\ldots,n-1$, denotes the determinant of the induced metric $\gamma_{ij}$ on the hypersurface $\partial V$, and $n^{\mu}$ are the coordinate components of the unit normal vector field to the boundary of $V$ denoted by $\partial V$.}\\

We now generalize the divergence theorem to the case in which the underlying Lorentzian manifold is endowed with a non-Riemannian geometry, namely, a Weyl--Integrable geometry. Given the compatibility condition \eqref{c3} the coordinate components of the Weyl affine connection coefficients read
\begin{equation}\label{eq26}
    \Gamma^{\alpha}_{\mu\nu}=\lbrace\,^{\alpha}_{\mu\nu}\rbrace -\frac{1}{2}g^{\alpha\beta}\left(g_{\beta\mu}\partial_{\nu}\varphi+g_{\beta\nu}\partial_{\mu}\varphi -g_{\mu\nu}\partial_{,\beta}\varphi\right),
\end{equation}
where the first term on the right hand side corresponds to the coordinate components of the Levi-Civita connection (also known as  Christoffel symbols of the second type). By definition, the divergence of the vector field $A^{\mu}$ is given by
\begin{equation}\label{eq27}
    \nabla_{\mu}A^{\mu}
    =
    \partial_{\mu}A^{\mu}
    +
    \Gamma^{\mu}_{\mu\lambda}A^{\lambda}.
\end{equation}
Now, with the help of \eqref{eq26} the contracted affine connection coefficient is
\begin{equation}
    \Gamma^{\mu}_{\mu\lambda}
    =
    \left\{
    \,^{\mu}_{\mu\lambda}
    \right\}
    -\frac{n}{2}\partial_{\lambda}\varphi.
\label{eq28}
\end{equation}
Substituting \eqref{eq28} into \eqref{eq27}, we obtain
\begin{equation}\label{eq29}
    \nabla_{\mu}A^{\mu}
    =
    \partial_{\mu}A^{\mu}
    +
    \left\{
    \,^{\mu}_{\mu\lambda}
    \right\}
    A^{\lambda}
    -
    \frac{n}{2}\,
    \partial_{\lambda}\varphi\,A^{\lambda}.
\end{equation}
It is not difficult to verify that the equation \eqref{eq29} can be rewritten as
\begin{equation}\label{eq30}
    \nabla_{\mu}A^{\mu}
    =
    \frac{1}{\sqrt{-g}}
    \partial_{\mu}
    \left(
    \sqrt{-g}\,A^{\mu}
    \right)
    -
    \frac{n}{2}\,
    \partial_{\mu}\varphi\,A^{\mu}.
\end{equation}
For later convenience, the equation \eqref{eq30} may also be written as
\begin{equation}\label{eq31}
    \nabla_{\mu}A^{\mu}
    =
    \frac{1}{\sqrt{-g}}
    \partial_{\mu}
    \left(
    \sqrt{-g}\,A^{\mu}
    \right)
    -
    \frac{n}{2\sqrt{-g}}\,
    \sqrt{-g}\,
    \partial_{\mu}\varphi\,A^{\mu}.
\end{equation}
Multiplying and dividing the right-hand side of \eqref{eq31} by the factor $e^{-n\varphi/2}$, we obtain
\begin{equation}\label{eq32}
    \nabla_{\mu}A^{\mu}
    =
    \frac{1}
    {\sqrt{-g}\,e^{-n\varphi/2}}
    \left[
    e^{-n\varphi/2}
    \partial_{\mu}
    \left(
    \sqrt{-g}\,A^{\mu}
    \right)
    -
    \frac{n}{2}
    \sqrt{-g}\,
    e^{-n\varphi/2}
    \partial_{\mu}\varphi\,A^{\mu}
    \right].
\end{equation}
Thus, applying the Leibniz rule yields the compact expression
\begin{equation}\label{eq33}
    \nabla_{\mu}A^{\mu}
    =
    \frac{e^{n\varphi/2}}
    {\sqrt{-g}}
    \partial_{\mu}
    \left(
    \sqrt{-g}\,
    e^{-n\varphi/2}
    A^{\mu}
    \right).
\end{equation}
On the other hand, as we mentioned before, due to the Weyl transformation \eqref{wt1} the differential line element is not Weyl-invariant: $d\bar{s}^{2}=e^{f(x)}ds^2$. Thus in an n-dimensional space-time 
\begin{equation}\label{eq37}
    \sqrt{-\bar{g}}=e^{\frac{n}{2}f(x)}\,\sqrt{-g}\,.
\end{equation}
Consequently, although the volume element is invariant under the group of diffeomorphisms, it is not invariant under Weyl transformations. Under the same group of transformations  the volume element transforms according to
\begin{equation}\label{eq38}
    d\bar{V}
    =
    \sqrt{-\bar{g}}\,d^{n}x
    =
    e^{\frac{n}{2}f(x)}
    \sqrt{-g}\,d^{n}x
    =
    e^{\frac{n}{2}f(x)}\,dV.
\end{equation}
Therefore, in order to formulate physical quantities that are invariant under both the diffeomorphism and Weyl groups, it is natural to introduce a volume element that remains invariant under both symmetries. We thus define the invariant volume element by
\begin{equation}\label{eq39}
    dV_{I}
    =
    e^{-\frac{n}{2}\varphi}
    \sqrt{-g}\,
    d^{n}x.
\end{equation}
It is straightforward to verify that $dV_{I}$ is invariant under the Weyl transformations. Hence, the Gauss divergence theorem in Weyl-integrable geometry must be formulated in terms of the invariant volume element defined in \eqref{eq39}. Accordingly
\begin{equation}\label{eq41}
    \int_{V}
    dV_{I}\,
    \nabla_{\mu}A^{\mu}
    =
    \int_{V}
    d^{n}x\,
    e^{-\frac{n}{2}\varphi}
    \sqrt{-g}\,
    \nabla_{\mu}A^{\mu}.
\end{equation}
By employing \eqref{eq33} the equation \eqref{eq41} leads to 
\begin{equation}\label{eq42}
    \int_{V}d^{n}x\,e^{-\frac{n}{2}\varphi}\,\sqrt{-g}\,\,\nabla_{\mu}A^{\mu}=\int_{V}d^{n}x\,\partial_{\mu}(\sqrt{-g}\,e^{-\frac{n}{2}\varphi}A^{\mu})
\end{equation}
which with the help of the traditional divergence theorem 
\begin{equation}\label{eq6}
    \int_{V}d^{n}x\,\partial_{\mu}(\sqrt{-g}\,A^{\mu})=\oint_{\partial V} A^{\mu}\,\sqrt{-g}\,d\sigma_{\mu},
\end{equation}
being $d\sigma_{\mu}$ the differential element of surface in the hypersurface $\partial V$, allow us to finally state the extended version of the divergence theorem for a Weyl integrable geometry in the form:\\

{\bf Theorem:} {\it Let $(M,g)$ be an $n$-dimensional Lorentzian manifold endowed with a metric $g$ and an affine connection $\nabla$ satisfying the Weyl-integrable compatibility condition
\begin{equation}\label{eq23}
    \frac{Dg}{d\sigma}(\mathbf{u},\mathbf{v})
    =
    d\varphi
    \left(
    \frac{d}{d\sigma}
    \right)
    g(\mathbf{u},\mathbf{v}),
\end{equation}
where $\lambda(\sigma):I\subseteq\mathbb{R}\rightarrow M$ is a smooth curve and $\mathbf{u},\mathbf{v}\in T_{\lambda(\sigma)}M$. Let $\mathbf{A}=A^{\mu}\partial_{\mu}\in\chi(M)$ be a smooth vector field whose coordinate components are invariant under Weyl transformations. Then, in any coordinate chart, the following identity holds
\begin{equation}\label{eq24}
    \int_{V}
    d^{n}x\,
    e^{-n\varphi/2}
    \sqrt{-g}\,
    \nabla_{\mu}A^{\mu}
    =
    \oint_{\partial V}
    A^{\mu}n_{\mu}\,
    e^{-(n-1)\varphi/2}
    \sqrt{|\gamma|}\,
    d^{(n-1)}x.
\end{equation}
Here, $\gamma=\det(\gamma_{ij})$, with $i,j=1,\ldots,n-1$, denotes the determinant of the induced metric on the boundary hypersurface $\partial V$, and $n^{\mu}$ are the coordinate components of the outward-pointing unit normal vector to $\partial V$.}\\

Thus, for a 4D space-time the formula \eqref{eq24} reduces to 

\begin{equation}\label{eq24pc}
    \int_{V}
    d^{4}x\,
    e^{-2\varphi}
    \sqrt{-g}\,
    \nabla_{\mu}A^{\mu}
    =
    \oint_{\partial V}
    A^{\mu}n_{\mu}\,
    e^{-3\varphi/2}
    \sqrt{|\gamma|}\,
    d^{3}x.
\end{equation}
However, we would like to bring the attention in an important point in \eqref{eq24} and \eqref{eq24pc}. If we start considering that under Weyl transformations $\bar{A}_{\mu}=A_{\mu}$ then as $A^{\alpha}=g^{\alpha\beta}A_{\beta}$, when we apply the Weyl transformation \eqref{wt1} we arrive to
\begin{equation}\label{at}
    \bar{A}^{\mu}=\bar{g}^{\mu\nu}\bar{A}_{\nu}=e^{-f}g^{\mu\nu}A_{\nu}=e^{-f}A^{\mu}.
\end{equation}
This expression is suggesting that in the Weyl-integrable geometry even if $A_{\mu}$ is Weyl-invariant its contravariant part $A^{\mu}$ is not. Consequently the term $\nabla_{\mu}A^{\mu}$ in \eqref{eq24} and \eqref{eq24pc} is not Weyl-invariant. However, the combination $e^{\varphi}A^{\mu}$ transforms according to
\begin{equation}\label{a20}
    e^{\bar{\varphi}}\bar{A}^{\mu}
    =
    e^{\varphi}
    e^{f}
    A^{\mu}
    e^{-f}
    =
    e^{\varphi}A^{\mu},
\end{equation}
showing that $e^{\varphi}A^{\mu}$ is invariant under Weyl transformations. Consequently, in order for the integrand on the left-hand side of \eqref{eq24} to be Weyl invariant, the replacement
\begin{equation}\label{a21}
    A^{\mu}
    \longrightarrow
    e^{\varphi}A^{\mu}
\end{equation}
must be performed. Accordingly, the equation \eqref{eq24} takes the form
\begin{equation}\label{a22}
    \int_{V}
    d^{n}x\,
    e^{-n\varphi/2}
    \sqrt{-g}\,
    \nabla_{\mu}
    \left(
    e^{\varphi}A^{\mu}
    \right)
    =
    \oint_{\partial V}
    A^{\mu}n_{\mu}\,
    e^{-(n-3)\varphi/2}
    \sqrt{|\gamma|}
    \,d^{(n-1)}x.
\end{equation}
This result provides a Weyl-invariant extension of the Gauss divergence theorem. In 4D this expression reduces to
\begin{equation}\label{a22a}
    \int_{V}
    d^{4}x\,
    e^{-2\varphi}
    \sqrt{-g}\,
    \nabla_{\mu}
    \left(
    e^{\varphi}A^{\mu}
    \right)
    =
    \oint_{\partial V}
    A^{\mu}n_{\mu}\,
    e^{-\varphi/2}
    \sqrt{|\gamma|}
    \,d^{3}x.
\end{equation}

On the other hand, the equation \eqref{a22a} can be recast into a form that is more suitable for the variational procedure used to obtain the field equations derived from \eqref{c10}. \\

With this in mind, let us start  calculating the quantity 
 $\nabla^{\mu}A_{\mu}$. Thus, by definition
\begin{equation}
    \nabla^{\mu}A_{\mu} = g^{\mu\sigma}\nabla_{\sigma}A_{\mu}= g^{\mu\sigma}\partial_{\sigma}A_{\mu}-g^{\mu\sigma}\lbrace\,^{\lambda}_{\sigma\mu}\rbrace A_{\lambda}-g^{\mu\sigma}W^{\lambda}_{\sigma\mu}A_{\lambda},\label{a23}
\end{equation}
where
\begin{equation}\label{a24}
    W^{\lambda}_{\sigma\mu}= \frac{1}{2}\left(\partial_{\mu}\varphi \delta^{\lambda}_{\sigma}+\partial_{\sigma}\varphi\delta^{\lambda}_{\mu}-g^{\lambda\beta}g_{\sigma\mu}\partial_{\beta}\varphi\right).
\end{equation}
With the help of \eqref{a24} a simplification of \eqref{a23} leads to
\begin{equation}\label{a25}
   \nabla^{\mu}A_{\mu} = g^{\mu\sigma}\partial_{\sigma}A_{\mu}-g^{\mu\sigma}\lbrace\,^{\lambda}_{\sigma\mu}\rbrace A_{\lambda}+\left(1-\frac{n}{2}\right)\partial_{\mu}A^{\mu}.
\end{equation}
Now, employing the standard formula 
\begin{equation}\label{a26}
g^{\sigma\mu}\lbrace\,^{\lambda}_{\sigma\mu}\rbrace =-\frac{1}{\sqrt{-g}} \,\partial_{\mu}(\sqrt{-g}g^{\lambda\mu}),
\end{equation}
the expression \eqref{a25} becomes
\begin{equation}\label{a27}
    \nabla^{\mu}A_{\mu} = g^{\mu\sigma}\partial_{\sigma}A_{\mu}+ \frac{1}{\sqrt{-g}} \,\partial_{\mu}(\sqrt{-g}g^{\lambda\mu})+\left(1-\frac{n}{2}\right)\partial_{\mu}A^{\mu}.
\end{equation}
After a little algebra the equation \eqref{a27} can be rewritten in the form
\begin{equation}
    \nabla^{\mu}A_{\mu} 
    = \frac{e^{\varphi}}{\sqrt{-g}}\left[\sqrt{-g}e^{-\varphi}g^{\mu \sigma}\partial_{\sigma}A_{\mu}+e^{-\varphi}\partial_{\sigma}\left(\sqrt{-g}\, g^{\mu\sigma}\right)A_{\mu}+\left(1-\frac{n}{2}\right)\sqrt{-g}\partial_{\sigma}\varphi \,g^{\mu\sigma}A_{\mu}\right],\label{a28}
\end{equation}
which for $n=4$ reads
\begin{equation}\label{a29}
  \nabla^{\mu}A_{\mu} =   \frac{e^{\varphi}}{\sqrt{-g}}\left[\sqrt{-g}e^{-\varphi}g^{\mu \sigma}\partial_{\sigma}A_{\mu}+e^{-\varphi}\partial_{\sigma}\left(\sqrt{-g}\, g^{\mu\sigma}\right)A_{\mu}+(\partial_{\sigma}e^{-\varphi})\sqrt{-g}\,g^{\mu\sigma}A_{\mu}\right].
\end{equation}
Simplifying \eqref{a29} we obtain
\begin{equation}\label{a30}
     \nabla^{\mu}A_{\mu} =\frac{e^{\varphi}}{\sqrt{-g}}\partial_{\mu}\left(\sqrt{-g}\,e^{-\varphi}A^{\mu}\right).
\end{equation}
Hence, with the help of \eqref{eq33} for the case $n=4$ and \eqref{a30} we arrive to the relation 
\begin{equation}\label{a31}
    e^{-\varphi}\nabla_{\mu}\left(e^{\varphi }A^{\mu}\right)=\nabla^{\mu}A_{\mu}.
\end{equation}
Therefore, employing \eqref{a31} the formula \eqref{a22a} can be recast in the form
\begin{equation}\label{a32}
    \int_{V} d^{4}x\,e^{-\varphi}\,\sqrt{-g}\,\,\nabla^{\mu}A_{\mu}=\oint_{\partial V} A^{\mu}n_{\mu}\,e^{-\frac{\varphi}{2}}\,\sqrt{|\gamma|}\,d^{3}x\,.
\end{equation}
This expression will be crucial in the moment we apply boundary conditions during the variational process to calculate the field equations from \eqref{c10}.

\section{The Weyl-invariant field equations}

We are now in position to obtain the field equations from the action \eqref{c10}. We begin by evaluating the variation of the action with respect to the metric tensor. To this end, the action \eqref{c10} is first recast in the form
\begin{equation}\label{c16}
     S=\int d^4x\,\sqrt{-g}\,e^{-\varphi}\left[\frac{1}{2\kappa} R-\frac{1}{2}\omega(\varphi)\,g^{\mu\nu}Y_{\mu}Y_{\nu}+e^{-\varphi}U(\varphi)-\frac{1}{4}e^{\varphi}g^{\alpha\mu}g^{\beta\nu}\Psi_{\alpha\beta}\Psi_{\mu\nu}\right],
\end{equation}
where
\[
Y_{\mu}={\cal D}_{\mu}\varphi=\partial_{\mu}\varphi+\gamma B_{\mu}\varphi.
\]
Thus, the variation of \eqref{c16} with respect to $g_{\mu\nu}$ leads to
\begin{eqnarray}
    \delta_{g}S 
     &=&\int d^{4}x\,e^{-\varphi}\sqrt{-g}\,\left[\frac{1}{2\kappa}\,{\cal G}_{\mu\nu}-\frac{1}{2} \,\left(\omega(\varphi)Y_{\mu}Y_{\nu}-\frac{1}{2}\omega(\varphi)g_{\mu\nu}Y_{\sigma}Y^{\sigma}\right)\right. \nonumber\\
     && \left.-\frac{1}{2}\,g_{\mu\nu}\,e^{-\varphi}U(\varphi)-\frac{1}{4}\,e^{\varphi}\left(-\frac{1}{2}\,g_{\mu\nu}\,\Psi^{\alpha\beta}\Psi_{\alpha\beta}+2\Psi_{\mu\alpha}\Psi_{\nu}\,^{\alpha}\right)\right]\,\delta g^{\mu\nu}, \nonumber\\
     && +\int d^4x\,e^{-\varphi}\,\nabla_{\mu}{\cal A}^{\mu}=0
     \label{c28}
\end{eqnarray}
where 
\begin{equation}\label{c28a}
    {\cal A}^{\mu}=g^{\rho\nu}\delta\Gamma^{\mu}_{\nu\rho}-g^{\rho\mu}\delta\Gamma^{\nu}_{\nu\rho}.
\end{equation}
It is worth emphasizing that the last term in \eqref{c28} must be treated with care. If one employs the standard divergence theorem, in which the integration measure does not contain the factor $e^{-\varphi}$, the exponential factor must first be absorbed into the co\-va\-riant divergence in order to apply the usual boundary condition that ${\cal A}^{\mu}$ vanishes on the hypersurface at infinity. This procedure generates additional contributions in the va\-ria\-tio\-nal principle and, consequently, modifies the resulting field equations. By contrast, within the present Weyl-invariant formulation, the appropriate divergence theorem requires the factor $e^{-\varphi}$ to remain outside the divergence operator, so that no extra terms arise after applying the boundary conditions. Therefore, the explicit form of the field equations depends on the geometric version of the divergence theorem adopted, leading, in principle, to different gravitational dynamics and hence to distinct physical scenarios. \\

Thus, imposing the boundary condition $\delta\Gamma^{\mu}_{\nu\alpha}\rightarrow 0$
on the boundary hypersurface $\partial V$, which is subsequently taken to infinity, with the use of \eqref{a32} the equation  \eqref{c28} implies 
\begin{equation}\label{c29}
    {\cal G}_{\mu\nu}=\kappa\left[\omega(\varphi)T_{\mu\nu}^{(Y)}+g_{\mu\nu}e^{-\varphi}U(\varphi)+e^{\varphi}T_{\mu\nu}^{(\Psi)}\right]\,,
    \end{equation}
being ${\cal G}_{\mu\nu}\equiv R_{\mu\nu}-\frac{1}{2}g_{\mu\nu}R=0$  the Einstein tensor where both $R_{\mu\nu}$ and $R$ are calculated from the Weyl affine connection coefficients \eqref{eq26} and 
\begin{eqnarray}
    T_{\mu\nu}^{(Y)}&=& Y_{\mu}Y_{\nu}-\frac{1}{2}g_{\mu\nu}Y_{\alpha}Y^{\alpha},\label{c30}\\
    T_{\mu\nu}^{(\Psi)}&=& \Psi_{\mu\alpha}\Psi_{\nu}\,^{\alpha}-\frac{1}{4}g_{\mu\nu}\Psi^{\alpha\beta}\Psi_{\alpha\beta}.\label{c31}
\end{eqnarray}
The equations \eqref{c29} are the field equations corresponding to the variation with respect to the metric tensor. \\

Now, the variation of \eqref{c16} with respect to the Weyl scalar field $\varphi$ take us to the expression
\begin{eqnarray}
    \delta_{\varphi}S &=& \int d^4x\,\sqrt{-g}\,e^{-\varphi}\left\lbrace -\frac{R}{2\kappa}+\frac{1}{2}\omega(\varphi)Y_{\mu}Y^{\mu}+\frac{1}{2}\omega^{\prime}(\varphi)Y_{\mu}Y^{\mu}-\gamma\omega^{\prime}(\varphi)\varphi B^{\mu}Y_{\mu}\right.\nonumber\\
    &+&\omega(\varphi)\,^{(w)}\Box\varphi+\gamma\omega(\varphi)\left(\nabla^{\mu}(\varphi B_{\mu})-B^{\mu}Y_{\mu}\right)+e^{-\varphi}\left(U^{\prime}(\varphi)-2U(\varphi)\right)\nonumber\\
    &-& \left. e^{\varphi}\left[(\nabla^{\nu}\varphi)\psi^{\mu}\,_{\nu}+\nabla^{\nu}\psi^{\mu}\,_{\nu}\right]B_{\mu}\right\rbrace \delta\varphi + \int \sqrt{-g}\,e^{-\varphi}\,\nabla^{\mu}\left(\frac{1}{2}\omega(\varphi)Y_{\mu}\delta\varphi\right)\nonumber\\
    &-& \int \sqrt{-g}\,e^{-\varphi}\,\nabla^{\nu}\left(e^{\varphi}\psi^{\mu}\,_{\nu} B_{\mu}\delta\varphi\right) = 0\,.
    \label{c56}
\end{eqnarray}
Again, similarly to the previous case, imposing the boundary condition $\delta\varphi \longrightarrow 0$ on the boundary hypersurface $\partial V \longrightarrow \infty$, the equation \eqref{a32} reduces \eqref{c56} to 
\begin{eqnarray}
&-&\frac{R}{2\kappa}+\frac{1}{2}\omega(\varphi)Y_{\mu}Y^{\mu}+\frac{1}{2}\omega^{\prime}(\varphi)Y_{\mu}Y^{\mu}
+\omega(\varphi)\,^{(w)}\Box\varphi+e^{-\varphi}\left(U^{\prime}(\varphi)-2U(\varphi)\right)-
\nonumber\\
&&\gamma\omega^{\prime}(\varphi)\varphi B^{\mu}Y_{\mu}+\gamma\omega(\varphi)\left(\nabla^{\mu}(\varphi B_{\mu})-B^{\mu}Y_{\mu}\right)-e^{\varphi}\left[(\nabla^{\nu}\varphi)\psi^{\mu}\,_{\nu}+\nabla^{\nu}\psi^{\mu}\,_{\nu}\right]B_{\mu}=0\,,\nonumber\\
    \label{c57} 
\end{eqnarray}
where $\,^{(w)}\Box\varphi$ is the D'Almabertian operator calculated with the Weyl affine connection coefficients  \eqref{eq26}. \\

On the other hand, to eliminate the explicit dependence on the Ricci scalar $R$ in the equation \eqref{c57}, we take the trace of the field equation \eqref{c29}. This process yields
\begin{equation}\label{c58}
    -R=\kappa\omega(\varphi)T^{(Y)}+4\kappa\,e^{-\varphi}\,U(\varphi)\,,
\end{equation}
where we have used the fact that the trace of the energy--momentum tensor $T^{(\psi)}_{\alpha\beta}$ va\-nishes. Moreover,
\begin{equation}\label{c59}
    T^{(Y)}=g^{\mu\nu}Y_{\mu}Y_{\nu}-\frac{1}{2}g^{\mu\nu}g_{\mu\nu}Y_{\alpha}Y^{\alpha}
    =-Y_{\alpha}Y^{\alpha}\,.
\end{equation}
Using \eqref{c59}, the expression \eqref{c58} leads to
\begin{equation}\label{c60}
R=\kappa\omega(\varphi)Y_{\mu}Y^{\mu}-4\kappa\,e^{-\varphi}U(\varphi)\,.
\end{equation}
Substituting \eqref{c60} in the first term of \eqref{c57}, we obtain
\begin{equation}\label{c61}
    -\frac{R}{2\kappa}
    =-\frac{1}{2}\omega(\varphi)Y_{\mu}Y^{\mu}
    +2e^{-\varphi}U(\varphi)\,.
\end{equation}
Finally, substituting \eqref{c61} in \eqref{c57} and simplifying, we arrive to
\begin{align}\label{c62}
    \begin{aligned}
    &\omega(\varphi)\,^{(w)}\Box\varphi+\frac{1}{2}\omega^{\prime}(\varphi)Y_{\mu}Y^{\mu}+e^{-\varphi}U^{\prime}(\varphi)-\gamma\omega^{\prime}(\varphi)\varphi B^{\mu}Y_{\mu}\\
    & +\gamma\omega(\varphi)\left[\nabla^{\mu}(\varphi B_{\mu})-B^{\mu}Y_{\mu}\right]-e^{\varphi}\left(\psi^{\mu}\,_{\nu}\nabla^{\nu}\varphi+\nabla^{\nu}\psi^{\mu}\,_{\nu}\right)=0\,.
    \end{aligned}
\end{align}
This is the field equation that determines the dynamics of the Weyl scalar field $\varphi$.\\

We can now  calculate the variation of \eqref{c16} with respect to the Weyl gauge field $B_{\lambda}$. Implementing then the variation we arrive to 
\begin{eqnarray}
    \delta_{B}S &=&\int d^4 x \sqrt{-g} e^{-\varphi}\left[-\omega(\varphi)(\gamma D^{\nu}\varphi)+\partial_{\mu}\varphi\,e^{\varphi}\psi^{\mu\nu}+\nabla_{\mu}(e^{\varphi}\psi^{\mu\nu})\right]\varphi\delta B_{\nu} \nonumber\\
    &-& \int d^{4}x\sqrt{-g}\,e^{-\varphi}\nabla^{\mu}(e^{\varphi}\psi_{\mu}\,^{\nu}\varphi\delta B_{\nu})\,.
    \label{c72}
\end{eqnarray}
The last term in the expression \eqref{c72} is a boundary contribution and, under the assumed boundary conditions, does not contribute to the variation of the action. Therefore, the equation \eqref{c72} reduces to
\begin{equation}\label{c73}
     \delta_{B}S =\int d^4 x \sqrt{-g} e^{-\varphi}\left[-\omega(\varphi)(\gamma D^{\nu}\varphi)+\partial_{\mu}\varphi\,e^{\varphi}\psi^{\mu\nu}+\nabla_{\mu}(e^{\varphi}\psi^{\mu\nu})\right]\varphi\delta B_{\nu}=0\,.
\end{equation}
Since the variation $\delta B_{\nu}$ is arbitrary, the corresponding Euler-Lagrange equation coming from \eqref{c73} is
\begin{equation}
 \nabla_{\mu}(e^{\varphi}\psi^{\mu\nu})+(\nabla_{\mu}e^{\varphi})\psi^{\mu\nu}
 = \gamma \omega(\varphi) D^{\nu}\varphi\,.
\end{equation}
This expression represents the field equation for the Weyl gauge field $B_{\nu}$.\\

\section{The field equations in the Einstein-Riemann frame}

In this section we will focus on figuring out how to obtain a purely relativistic theory, that is, with a Riemannian background geometry. This can be done through the introduction of the named Einstein-Riemann frame. As we will show below, this frame is characterized because the background geometry is effective  Riemannian. 
The transition from the Weyl frame to the Einstein--Riemann frame is particularly useful for clarifying the physical content of scalar--tensor theories with non-minimal coupling in a Weyl Integrable geometry. Although the Weyl frame provides a natural geometric description in which the scalar field is intrinsically related to the non-metricity of the spacetime, the corresponding gravitational sector is generally non-canonical. The idea is that using the Weyl group of transformation and a metric redefinition   the theory can be recast in the Einstein--Riemann frame, where the gravitational action takes the standard Einstein--Hilbert form and the effects of the non-minimal coupling are transferred to the scalar sector. This formulation facilitates the identification of the effective scalar dynamics and a\-llows possible cosmological predictions of the model to be interpreted in terms of the familiar Riemannian framework.\\

Let us begin with the definition of an Einstein-Riemann frame. 
To derive the field equations in the Einstein--Riemann frame, we first note that, for the gauge choice $f=-\varphi$, the Weyl transformations \eqref{wt1} and \eqref{wt2} reduce to
\begin{eqnarray}
    \bar{g}_{\mu\nu} &=& e^{-\varphi}\,g_{\mu\nu}\,, \label{erf1}\\
    \bar{\varphi} &=& 0\,.
    \label{erf2}
\end{eqnarray}
Recall that, in general, Weyl transformations map one Weyl-integrable frame $(M,g,\varphi)$ into another frame $(M,\bar{g},\bar{\varphi})$, preserving the underlying Weyl-integrable geometric structure. For the particular gauge choice $\bar{\varphi}=0$, the action \eqref{c16} becomes
\begin{equation}\label{erf3}
    \bar{S}(\bar{g})=\int d^4 x\,\sqrt{-\bar{g}}\,\frac{\bar{R}}{2\kappa}\,,
\end{equation}
which has the form of the Einstein--Hilbert action, where the Ricci scalar is constructed from the effective affine connection
\begin{equation}\label{erf4}
    \Gamma^{\alpha}_{\mu\nu}=
    \frac{1}{2}\bar{g}^{\alpha\sigma}
    \left(
    \partial_{\nu}\bar{g}_{\sigma\mu}
    +\partial_{\mu}\bar{g}_{\sigma\nu}
    -\partial_{\sigma}\bar{g}_{\mu\nu}
    \right).
\end{equation}
It is evident that the equation \eqref{erf4} is precisely the Levi--Civita connection associated with the metric $\bar{g}_{\mu\nu}$. Consequently, the metric compatibility condition associated to $\bar{g}_{\alpha\beta}$ takes the form
\begin{equation}\label{erf5}
    \nabla_{\alpha}\bar{g}_{\mu\nu}
    =
    \partial_{\alpha}\bar{\varphi}\,
    \bar{g}_{\mu\nu}
    =
    0,
\end{equation}
which is the standard metricity condition of Riemannian geometry. Therefore, the frame $(M,\bar{g},\bar{\varphi}=0)$ corresponds to an effective Riemannian geometry, and the vacuum sector of general relativity is recovered. For this reason, it is convenient to identify this frame simply as $(M,\bar{g})$.\\

However, it is important to emphasize that the gauge choice $\bar{\varphi}=0$ eliminates the Weyl scalar field from the geometric sector of the theory, so that gravitation is entirely described by the metric tensor in this gauge. Nevertheless, this gauge-fixed should not be confused with the Einstein--Riemann frame introduced below. The latter is not obtained by imposing an additional gauge condition, but rather by reformulating the theory in terms of the Weyl-invariant effective metric
\begin{equation}\label{erf6}
    h_{\mu\nu}
    =
    e^{-\varphi}g_{\mu\nu}
    =
    \bar{g}_{\mu\nu}.
\end{equation}
The Einstein--Riemann frame is therefore defined as the Riemannian manifold $(M,h)$, whose metric satisfies
\begin{eqnarray}
\nabla_{\alpha}h_{\mu\nu}
&=&
\nabla_{\alpha}
\left(
e^{-\varphi}g_{\mu\nu}
\right)
\nonumber\\
&=&
-\partial_{\alpha}\varphi\,
e^{-\varphi}g_{\mu\nu}
+
e^{-\varphi}
\nabla_{\alpha}g_{\mu\nu}
\nonumber\\
&=&
-\partial_{\alpha}\varphi\,
e^{-\varphi}g_{\mu\nu}
+
\partial_{\alpha}\varphi\,
e^{-\varphi}g_{\mu\nu}
\nonumber\\
&=&
0,
\label{erf7}
\end{eqnarray}
where the Weyl non-metricity condition
$\nabla_{\alpha}g_{\mu\nu}
=
\partial_{\alpha}\varphi\,g_{\mu\nu}$
has been used. Consequently, the affine connection associated with $h_{\mu\nu}$ is the Levi--Civita connection, and the geo\-metry of $(M,h)$ is purely Riemannian.\\

The Einstein--Riemann frame retains the scalar field $\varphi$ as an independent dynamical degree of freedom. Since the geometry is now Riemannian, the scalar field no longer enters the affine connection and therefore ceases to play a geometrical role. Nevertheless, because it originates from the Weyl geometry of the original frame $(M,g,\varphi)$, it is naturally interpreted as a physical scalar field of geometric origin. In this sense, the transition from $(M,g,\varphi)$ to $(M,h)$ transforms $\varphi$ from a geometrical field into a physical one, while the gravitational sector is described by the metric. In the original Weyl-frame $(M,g,\varphi)$, as the affine connection is depending on both $g_{\mu\nu}$ and $\varphi$, then we can interpret this as gravity has two sectors in this frame: the tensor described by $g_{\mu\nu}$ and the scalar defined by $\varphi$.\\

Now, to rewrite the action \eqref{c16} in the Einstein--Riemann frame, we make use of the identities
\begin{eqnarray}
   \sqrt{-g} &=& e^{-2\varphi}\sqrt{-h}\,, \label{erf8}\\
   g_{\alpha\beta} &=& e^{\varphi}h_{\alpha\beta}\,,\label{erf9}\\
   g^{\alpha\beta} &=& e^{-\varphi} h^{\alpha\beta}\,, \label{erf10}\\
   R(g) &=& e^{-\varphi} R(h)\,, \label{erf11}\\
   \psi_{\mu\nu}\psi^{\mu\nu}
   &=&
   e^{-2\varphi}
   \psi_{\mu\nu}\psi^{\mu\nu}\,,
   \label{erf12}
\end{eqnarray}
where, in the last the indices on the right-hand side are raised with the metric $h_{\mu\nu}$. Substituting the expressions \eqref{erf8}-\eqref{erf12} into the action \eqref{c16} yields
\begin{equation}\label{erf13}
    S(h)=\int d^{4}x\,\sqrt{-h}\left[\frac{R(h)}{2\kappa}-\frac{1}{2}\omega(\varphi)h^{\mu\nu}{\cal D}_{\mu}\varphi{\cal D}_{\nu}\varphi+U(\varphi)-\frac{1}{4}\psi_{\mu\nu}\psi^{\mu\nu}\right].
\end{equation}
This is the corresponding form of the Weyl-invariant action \eqref{c10} in the Einstein--Riemann frame, where the gravitational sector is described by the Riemannian metric $h_{\mu\nu}$.\\

 We are now in a position to derive the field equations associated with the action \eqref{erf13}. It is important to remember that in this frame the background geometry is Riemannian, and thus the variational procedure is the usual. The variation of \eqref{erf13} with respect to the metric $h_{\alpha\beta}$ is given by
 \begin{eqnarray}
\delta_{h}S(h)&=&\int d^4x\,\sqrt{-h}\,\left[\frac{1}{2\kappa}G_{\mu\nu}-\frac{1}{2}\omega(\varphi)\left({\cal D}_{\mu}\varphi{\cal D}_{\nu}\varphi-\frac{1}{2}h_{\mu\nu}{\cal D}_{\alpha}\varphi{\cal D}^{\alpha}\varphi\right)-\frac{1}{2}U(\varphi)h_{\mu\nu}\right.\nonumber\\
&-&\left.\frac{1}{2}\left(-\frac{1}{4}\psi_{\alpha\beta}\psi^{\alpha\beta}+\psi_{\mu\alpha}\psi_{\nu}\,^{\alpha}\right)\right]\delta h^{\mu\nu}=0\,,
    \label{erf26}
\end{eqnarray}
being $G_{\mu\nu}=R_{\mu\nu}-\frac{1}{2}R g_{\mu\nu}$  the Einstein tensor calculated with the Levi-Civita connection for the metric $h_{\alpha\beta}$. It follows from \eqref{erf26} that
\begin{equation}\label{erf27}
 G_{\mu\nu}=\kappa \left(T_{\mu\nu}^{(\varphi)}+T_{\mu\nu}^{(B)}\right)\,,
\end{equation} 
where
\begin{eqnarray}
    T_{\mu\nu}^{(\varphi)}&=&\omega(\varphi){\cal D}_{\mu}\varphi{\cal D}_{\nu}\varphi-\frac{1}{2}h_{\mu\nu}\left(\omega(\varphi){\cal D}_{\alpha}\varphi{\cal D}^{\alpha}\varphi -2U(\varphi)\right)\,,
    \label{erf28}\\
    T_{\mu\nu}^{(B)}&=& \psi_{\mu\alpha}\psi_{\nu}\,^{\alpha}-\frac{1}{4}h_{\mu\nu}\psi_{\alpha\beta}\psi^{\alpha\beta}\,.
    \label{erf29}
\end{eqnarray}
These are the field equations that determine the dynamics of the metric tensor $h_{\mu\nu}$. \\

Now, the variation of \eqref{erf13} with respect to the scalar field $\varphi$ leads us to
\begin{eqnarray}
    \delta_{\varphi}S &=&\int d^4x\,\sqrt{-h}\left[\frac{1}{2}\omega^{\prime}(\varphi){\cal D}^{\mu}\varphi{\cal D}_{\mu}\varphi-\gamma \omega^{\prime}(\varphi)(\varphi B^{\mu}){\cal D}_{\mu}\varphi+\omega(\varphi)\Box\varphi\right.\nonumber\\
    &&\left. + \gamma \omega(\varphi)\left(\nabla^{\mu}(\varphi B_{\mu})-B^{\mu}{\cal D}_{\mu}\varphi\right)+U^{\prime}(\varphi)+(\nabla_{\mu}\psi^{\mu\nu})B_{\nu}\right]\delta\varphi=0\,,
    \label{qcg1}
\end{eqnarray}
which implies
\begin{eqnarray}
    && \omega(\varphi)\Box\varphi+\frac{1}{2}\omega^{\prime}(\varphi){\cal D}^{\mu}\varphi{\cal D}_{\mu}\varphi-\gamma \omega^{\prime}(\varphi)(\varphi B^{\mu}){\cal D}_{\mu}\varphi + \gamma\omega(\varphi)\left(\nabla^{\mu}(\varphi B_{\mu})-B^{\mu}{\cal D}_{\mu}\varphi\right)\nonumber\\
    && +U^{\prime}(\varphi)+(\nabla_{\mu}\psi^{\mu\nu})B_{\nu}=0\,,
    \label{qcg2}
\end{eqnarray}
where here $\Box =h^{\mu\nu}\nabla_{\mu}\nabla_{\nu}$ is the D'Alambertian operator calculated with the metric tensor $h_{\alpha\beta}$. \\

Similarly, the variation of the action \eqref{erf13} with respect to the gauge field $B_{\mu}$ take us to
\begin{equation}\label{qcg3}
    \delta_{B}S=\int d^4x\,\sqrt{-h}\left[\nabla_{\mu}\psi^{\mu\nu}-\gamma\omega(\varphi)\, {\cal D}^{\nu}\varphi\right]\varphi \,\delta B_{\nu}=0\,.
\end{equation}
Hence, it becomes from \eqref{qcg3} that
\begin{equation}\label{qcg4}
    \nabla_{\mu}\psi^{\mu\nu}=\gamma\omega(\varphi)\, {\cal D}^{\nu}\varphi\,.
\end{equation}
This is the equation that governs the dynamics of the Weyl gauge field $B_{\alpha}$. Therefore, the equations \eqref{erf27}, \eqref{qcg2} and \eqref{qcg4} are the field equations for the fields $h_{\mu\nu}$, $\varphi$ and $B_{\nu}$ in the Einstein-Riemann frame. \\

The equation \eqref{qcg4} can be written in terms of $\varphi B^{\mu}$ in the form
\begin{equation}\label{qcg5}
    \Box(\varphi B^{\nu})-\nabla_{\mu}\nabla^{\nu}(\varphi B^{\mu})=\gamma \omega(\varphi){\cal D}^{\nu}\varphi. 
\end{equation}
With the help of the relation 
\begin{equation}\label{qcg6}
    \left[\,\nabla_{\mu},\,\nabla_{\alpha}\right](\varphi B^{\mu})=R_{\lambda\alpha}(\varphi B^{\lambda})\,
\end{equation}
where $R_{\lambda\alpha}$ is the Ricci tensor calculated with the metric $h_{\alpha\beta}$, the equation \eqref{qcg5} can be recast as 
\begin{equation}\label{qcg7}
    \Box(\varphi B^{\nu})-\nabla^{\nu}(\nabla_{\mu}(\varphi B^{\mu}))-R^{\nu}\,_{\lambda}(\varphi B^{\lambda})=\gamma \omega(\varphi)(\nabla^{\nu}\varphi+\gamma \varphi B^{\nu})\,.
\end{equation}
This form could be more suitable to the study primordial quantum fluctuations in inflationary cosmological models.

\section{Explicit Geometrical breaking of Weyl symmetry}

The Weyl-invariant formulation introduced above provides a natural geometric framework in which the gravitational and matter sectors are described in terms of fields and geometric structures possessing a local scale symmetry. In particular, the Weyl transformations relate different representatives of the same conformal class of geometries, while the corresponding Weyl-invariant action remains unchanged under such transformations. This symmetry plays a central role in establishing the equivalence between the Weyl frame and the Einstein--Riemann frame.\\

In the present work, we consider a controlled departure from this symmetry by introducing an additional interaction between the Weyl gauge field and a matter current. Specifically, we extend the Weyl-invariant action according to
\begin{equation}
    S_{\rm W}
    \;\longrightarrow\;
    S_{\rm W}+\lambda\int d^4x\,\sqrt{-g}\,
    B_{\mu}J^{\mu},
    \label{SBcurrent}
\end{equation}
where $\lambda$ is a coupling constant, $B_{\mu}$ denotes the Weyl gauge field, and $J^{\mu}(\varphi,\partial_{\nu}\varphi)$ is a current constructed from the fields of the theory.\\

The essential point is that the current $J^{\mu}$ is assumed to possess a transformation law such that the contraction $B_{\mu}J^{\mu}$ does not transform as a Weyl scalar density with the appropriate weight to compensate the transformation of the measure. Consequently, the additional interaction in Eq.~\eqref{SBcurrent} is not invariant under local Weyl transformations. The resulting theory therefore possesses a reduced symmetry: diffeomorphism covariance is preserved, whereas local Weyl invariance is explicitly broken by the current coupling.\\

This mechanism can be interpreted as a \emph{geometrical breaking of Weyl symmetry}. The terminology reflects the fact that the symmetry breaking is introduced through a direct coupling to the field $B_{\mu}$ that characterizes the Weyl geometry itself. In this sense, the breaking is not introduced merely as an external potential or through an arbitrary non-invariant term, but through an interaction involving the geometric field responsible for the Weyl structure of spacetime.\\

It is important to distinguish this construction from a spontaneous breaking of Weyl symmetry. Here, the symmetry is broken explicitly by the presence of the current interaction in the action. In particular, the parameter $\lambda$ controls the departure from the Weyl-invariant theory. The limit $ \lambda\rightarrow 0$ smoothly restores the original Weyl symmetry.\\

In the absence of the current interaction, the transformation to the Einstein--Riemann frame yields the standard Riemannian representation of the original Weyl-invariant theo\-ry. However, once the term in Eq.~\eqref{SBcurrent} is included, its transformed counterpart remains present in the Einstein--Riemann frame. Since $B_{\mu}J^{\mu}$ is a scalar under diffeomorphisms, the resulting interaction is fully compatible with general covariance in the Riemannian geometry. Thus, the passage to the Einstein--Riemann frame does not eliminate the symmetry-breaking interaction; rather, it translates the geometrical breaking of Weyl symmetry into an additional generally covariant interaction in the Riemannian description.

This observation also establishes an important relation between the symmetric and symmetry-broken theories. Denoting by $S_{\rm ER}^{(0)}$ the action obtained in the Einstein--Riemann frame \eqref{erf13} from the Weyl-invariant theory, and by $\Delta S_{\rm ER}$ the transformed current interaction, the complete action can be schematically written as
\begin{equation}
    S_{\rm ER}
    =
    S_{\rm ER}^{(0)}
    +
    \lambda\,\Delta S_{\rm ER}.
    \label{SERgeneral}
\end{equation}
Therefore, $S_{\rm ER}$ represents a more general theory than the Einstein--Riemann action obtained from the strictly Weyl-invariant theory. The latter is recovered as the special case
\begin{equation}
    \lambda=0,
    \qquad
    S_{\rm ER}\big|_{\lambda=0}
    =
    S_{\rm ER}^{(0)}.
\end{equation}

The advantage of this construction is that it provides a direct way of parametrizing deviations from Weyl symmetry while maintaining diffeomorphism invariance. The coupling $\lambda$ consequently measures the strength of the geometrical departure from the Weyl-invariant theory. In this framework, the Weyl-invariant model is not replaced by an unrelated Riemannian theory; instead, it appears as a distinguished limit of a broader generally covariant theory containing a controlled Weyl-symmetry-breaking interaction.\\

From this perspective, the Einstein--Riemann frame provides a useful description of the symmetry-broken theory, while the Weyl frame makes the geometric origin of the breaking explicit. The two descriptions therefore emphasize complementary aspects of the same construction: the former exhibits a generally covariant Riemannian theory with an additional interaction, whereas the latter reveals that this interaction originates from a direct coupling to the field defining the Weyl geometry. \\

In this manner, breaking explicitly the Weyl symmetry in the action \eqref{c10} we arrive to the new action
\begin{equation}\label{mc1}
    S=\int d^4x\,\sqrt{-g}\,e^{-\varphi}\left[\frac{1}{2\kappa} R-\frac{1}{2}\omega(\varphi)\,g^{\mu\nu}{\cal D}_{\mu}\varphi{\cal D}_{\nu}\varphi+e^{-\varphi}U(\varphi)-\frac{1}{4}e^{\varphi}\Psi_{\mu\nu}\Psi^{\mu\nu}+\lambda B_{\mu}J^{\mu}\right]\,.
\end{equation}
As mentioned above, the mechanism for breaking Weyl symmetry is based on in\-tro\-du\-cing the interaction term $\lambda B_{\mu}J^{\mu}$, where the current is, in general, a functional of the scalar field and its first derivatives,
\begin{equation}
    J^{\mu}=J^{\mu}\left(\varphi,\partial_{\nu}\varphi\right).
\end{equation}
The current $J^{\mu}$ is assumed to transform in such a way that the corresponding interaction does not remain invariant under local Weyl transformations \eqref{wt1} and \eqref{wt2}. In particular, under  Weyl transformations, one has
\begin{equation}
    \bar{\lambda}\,\bar{B}_{\mu}\bar{J}^{\mu}
    =
    \lambda B_{\mu}J^{\mu}
    +\Delta_{\rm W},
    \label{mc2}
\end{equation}
where $\Delta_{\rm W}$ denotes the additional terms generated by the Weyl transformation, which do not vanish in general. Consequently, the presence of the current interaction ex\-pli\-ci\-tly breaks the Weyl symmetry of the original action, while preserving covariance under diffeomorphisms. \\

For example, in the case of $J^{\mu}=\gamma \varphi \omega(\varphi)\partial^{\mu}\varphi$, considering the expression \eqref{c13}, the term $\Delta_{\rm W}$ in the transformation \eqref{mc2} becomes
\begin{equation}\label{mc3}
    \Delta_{\rm W}=-\gamma^{-1}\lambda \omega(\varphi)\partial^{\mu}\varphi\partial_{\mu}f+\lambda\omega(\varphi)\partial^{\mu}f (\varphi B_{\mu}-\gamma^{-1}\partial_{\mu}f)\,.
\end{equation}
In the Einstein-Riemann frame the action \eqref{mc1} becomes
\begin{equation}\label{mc4}
    S(h)=\int d^{4}x\,\sqrt{-h}\left[\frac{R(h)}{2\kappa}-\frac{1}{2}\omega(\varphi)h^{\mu\nu}{\cal D}_{\mu}\varphi{\cal D}_{\nu}\varphi+\lambda B_{\mu}J^{\mu}+U(\varphi)-\frac{1}{4}\psi_{\mu\nu}\psi^{\mu\nu}\right]. 
\end{equation}
The field equations obtained from \eqref{mc4} then read
\begin{eqnarray}
  && G_{\mu\nu}=\kappa \left(T_{\mu\nu}^{(\varphi)}+T_{\mu\nu}^{(B)}+h_{\mu\nu}B_{\alpha}J^{\alpha}\right)\,,  \label{mc5}\\
  && \omega(\varphi)\Box\varphi+\frac{1}{2}\omega^{\prime}(\varphi){\cal D}^{\mu}\varphi{\cal D}_{\mu}\varphi-\gamma \omega^{\prime}(\varphi)(\varphi B^{\mu}){\cal D}_{\mu}\varphi + \gamma\omega(\varphi)\left(\nabla^{\mu}(\varphi B_{\mu})-B^{\mu}{\cal D}_{\mu}\varphi\right)\nonumber\\
    && +U^{\prime}(\varphi)+(\nabla_{\mu}\psi^{\mu\nu})B_{\nu}+\lambda B_{\mu}\left(\frac{\partial J^{\mu}}{\partial\varphi}-\partial_{\nu}\left(\frac{\partial J^{\mu}}{\partial (\partial_{\nu}\varphi)}\right)\right)=0\,, \label{mc6} \\
    && \nabla_{\mu}\psi^{\mu\nu}=\gamma\omega(\varphi){\cal D}^{\nu}\varphi-\lambda J^{\nu}\,, \label{mc7}
\end{eqnarray}
where $T_{\mu\nu}^{(\varphi)}$ and $T_{\mu\nu}^{(B)}$ are given respectively by \eqref{erf28} and \eqref{erf29}. 

\section{Concluding Remarks}

In this work, we have developed a geometrically motivated formulation of gravity in an invariant Weyl--Integrable space-time (IWIST), in which the background geometry is not imposed a priori but is dynamically determined through the Palatini variational principle. For a scalar field non-minimally coupled to gravity, the resulting compatibility condition naturally leads to a Weyl--Integrable geometry. This provides a geometrical basis for identifying the local Weyl transformations as the symmetry transformations associated with the underlying spacetime structure.

Following this principle, we have constructed a gravitational action invariant under both diffeomorphisms and local Weyl transformations. The Weyl scalar $\varphi$ is incorporated into the geometrical structure through the Weyl non-metricity condition, while the corresponding Weyl gauge field allows the construction of Weyl-covariant derivatives and field-strength tensors. In this way, the gravitational action, the underlying geo\-me\-try, and the symmetry transformations are consistently described within a common geometrical framework.

A further important ingredient of the formulation is the modification of the divergence theorem required by the Weyl geometry. This construction allows the boundary terms arising in the variational procedure to be treated consistently while preserving Weyl invariance. As a consequence, the field equations obtained from the invariant action possess the same geometrical symmetry as the action itself. The resulting equations describe the coupled dynamics of the metric, the Weyl scalar, and the Weyl gauge field.\\

We have subsequently reformulated the theory in the Einstein--Riemann frame by introducing the effective metric defined in \eqref{erf6}. In this frame, the affine connection becomes the Levi--Civita connection and the background geometry is purely Riemannian. Nevertheless, the scalar field $\varphi$ remains as an independent dynamical degree of freedom. It therefore changes its geometrical role: while it determines the non-metricity of the Weyl frame, in the Einstein--Riemann frame it is interpreted as a physical scalar field of geometric origin. The complete set of field equations in this frame is given by the equations \eqref{erf27}, \eqref{qcg2} and \eqref{qcg4}, providing a fully Riemannian representation of the original Weyl-invariant theory.\\

Finally, we have introduced a controlled departure from local Weyl invariance through the current interaction discussed in section 6. In particular, the additional coupling $\lambda B_{\mu}J^{\mu}$, with the current constructed in general from the scalar field and its first derivatives, is chosen so that it does not remain invariant under the Weyl transformations. As discussed in the equation \eqref{mc2}, this interaction explicitly breaks the local Weyl symmetry while preserving covariance under diffeomorphisms. We refer to this mechanism as a \emph{geometrical breaking of Weyl symmetry}, since the breaking is introduced through a direct coupling to the field that characterizes the Weyl geometry. Thus, the Weyl-invariant theory is recovered as the particular limit $\lambda=0$, whereas nonzero values of $\lambda$ describe a controlled departure from local Weyl invariance.\\

The resulting theory is more general than the strictly Weyl-invariant theory. In particular, the complete action obtained after introducing the symmetry-breaking interaction is given in \eqref{mc1}, while its Einstein--Riemann representation \eqref{mc4}  contains the corresponding additional interaction. The complete set of  field equations in the Einstein-Riemann frame for the symmetry-broken theory  are given by \eqref{mc5}-\eqref{mc7}. \\

An important aspect of this construction is that the breaking of Weyl symmetry does not require abandoning diffeomorphism covariance. Instead, the symmetry breaking is encoded in an additional interaction that has a clear geometrical origin in the Weyl frame and appears as a generally covariant contribution in the Einstein--Riemann description. The two frames therefore provide complementary perspectives on the same theory: the Weyl frame makes the geometrical origin of the symmetry and its breaking manifest, whereas the Einstein--Riemann frame provides a description in terms of a Riemannian geometry and physical fields.\\

The framework developed here provides a geometrically consistent setting in which Weyl invariance can be continuously deformed into a more general generally covariant theory. The coupling $\lambda$ consequently provides a natural parameter with which to characterize deviations from the Weyl-invariant theory. A detailed analysis of the cosmological and phenomenological consequences of this geometrical symmetry breaking, including its implications for the dynamics of the scalar sector, cosmological perturbations, and possible observational signatures, will be considered in future work.
	
	\section*{\large{Acknowledgements}}
	
	\noindent J. E. Madriz-Aguilar, A. Bernal,  M. Montes and J. Zamarripa acknowledge  SECIHTI
	(M\'exico) and Centro Universitario de Ciencias Exactas e Ingenierias of Universidad de Guadalajara for financial support.


\bigskip
	
	
	\begin{thebibliography}{99}


\bibitem{Weyl1918}
H. Weyl,
``Gravitation und Elektrizität,''
Sitzungsber. Preuss. Akad. Wiss. Berlin
(1918) 465--480.

\bibitem{Novello1983}
M. Novello,
``Weyl Integrable Space-Time: A Model of Our Cosmos?''
Phys. Lett. A {\bf 98}
(1983) 10--12.

\bibitem{Novello1992}
M. Novello, L. A. R. Oliveira, J. M. Salim and E. Elbaz,
``Geometrized Instantons and the Creation of the Universe,''
Int. J. Mod. Phys. D {\bf 1}
(1992) 641--677.

\bibitem{Oliveira1997}
H. P. Oliveira, S. M. Salim and S. E. Perez Bergliaffa,
``Non-singular Inflationary Cosmologies in Weyl Integrable Space-Time,''
Class. Quantum Grav. {\bf 14}
(1997) 2833--2843.

\bibitem{Romero2012}
C. Romero, J. B. Fonseca-Neto and M. L. Pucheu,
``General Relativity and Weyl Geometry,''
Class. Quantum Grav. {\bf 29}
(2012) 155015.

\bibitem{RomeroCF2012}
C. Romero, J. B. Fonseca-Neto and M. L. Pucheu,
``Conformally Flat Space-Times and Weyl Frames,''
Found. Phys. {\bf 42}
(2012) 224--240.

\bibitem{Avalos2018}
R. Avalos, F. Dahia and C. Romero,
``A Note on the Problem of Proper Time in Weyl Space-Time,''
Int. J. Geom. Methods Mod. Phys. {\bf 15}
(2018) 1850036.

\bibitem{Lobo2015}
I. P. Lobo, A. B. Barreto and C. Romero,
``Space-Time Singularities in Weyl Manifolds,''
Eur. Phys. J. C {\bf 75}
(2015) 448.

\bibitem{Madriz2015}
J. E. Madriz Aguilar, C. Romero, J. B. Fonseca-Neto,
T. S. Almeida and J. B. Formiga,
``(2+1)-Dimensional Gravity in Weyl Integrable Space-Time,''
Class. Quantum Grav. {\bf 32}
(2015) 215003.

\bibitem{Miritzis2004}
J. Miritzis,
``Isotropic Cosmologies in Weyl Geometry,''
Class. Quantum Grav. {\bf 21}
(2004) 3043--3055.

\bibitem{Paliathanasis2020}
A. Paliathanasis, G. Leon and J. D. Barrow,
``Inhomogeneous Space-Times in Weyl Integrable Geometry with Matter Source,''
Eur. Phys. J. C {\bf 80}
(2020) 731.

\bibitem{Chatzidakis2022}
S. Chatzidakis, A. Giacomini, G. Leon,
A. Paliathanasis and S. Pan,
``Interacting Dark Energy in Curved FLRW Space-Time from Weyl Integrable Geometry,''
J. High Energy Astrophys. {\bf 36}
(2022) 141--151.

\bibitem{IQMA} I. Quiros, R. García-Salcedo, J. E. Madriz-Aguilar, T. Matos, ``The conformal transformation's controversy: what are we missing?, '' Gen. Rel. Grav. {\bf 45} (2) (2013) 489-518.


\bibitem{U1}
V.~Faraoni, E.~Gunzig and P.~Nardone,
``Conformal transformations in classical gravitational theories and in cosmology,''
Fundam.\ Phys.\ \textbf{20}, 121--151 (1999).
arXiv:gr-qc/9811047.

\bibitem{U2}
R.~Catena, M.~Pietroni and L.~Scarabello,
``Einstein and Jordan frames reconciled: A frame-invariant approach to scalar-tensor cosmology,''
Phys.\ Rev.\ D \textbf{76}, 084039 (2007).
doi:10.1103/PhysRevD.76.084039.

\bibitem{U3}
M.~Postma and M.~Volponi,
``Equivalence of the Einstein and Jordan frames,''
Phys.\ Rev.\ D \textbf{90}, 103516 (2014).
doi:10.1103/PhysRevD.90.103516.
arXiv:1407.6874.

\bibitem{U4}
J.~R.~Morris,
``Consistency of equations of motion in conformal frames,''
Phys.\ Rev.\ D \textbf{90}, 107501 (2014).
doi:10.1103/PhysRevD.90.107501.
arXiv:1411.1311.

\bibitem{U5}
L.~Järv, P.~Kuusk, M.~Saal and O.~Vilson,
``Invariant quantities in the scalar-tensor theories of gravitation,''
Phys.\ Rev.\ D \textbf{91}, 024041 (2015).
doi:10.1103/PhysRevD.91.024041.

\bibitem{U6}
V.~Faraoni,
``Illusions of general relativity in Brans-Dicke gravity,''
Phys.\ Rev.\ D \textbf{59}, 084021 (1999).
doi:10.1103/PhysRevD.59.084021.
arXiv:gr-qc/9902083.

\bibitem{U7}
G.~Gionti,
``Canonical analysis of Brans-Dicke theory addresses Hamiltonian inequivalence between the Jordan and Einstein frames,''
Phys.\ Rev.\ D \textbf{103}, 024022 (2021).
doi:10.1103/PhysRevD.103.024022.

\bibitem{U8}
S.~Capozziello, R.~de~Ritis and A.~A.~Marino,
``Some aspects of the cosmological conformal equivalence between
Jordan frame and Einstein frame,''
Class.\ Quantum Grav.\ \textbf{14}, 3243--3258 (1997).
doi:10.1088/0264-9381/14/11/010.


\bibitem{U8A}
I.~P.~Lobo,
``On the physical interpretation of non-metricity in Brans--Dicke gravity,''
Int.\ J.\ Geom.\ Meth.\ Mod.\ Phys.\ \textbf{15}, 1850138 (2018).
doi:10.1142/S0219887818501384.
arXiv:1610.05004 [gr-qc].

\bibitem{U8B}
A.~Delhom, I.~P.~Lobo, G.~J.~Olmo and C.~Romero,
``A generalized Weyl structure with arbitrary non-metricity,''
Eur.\ Phys.\ J.\ C \textbf{79}, 878 (2019).
doi:10.1140/epjc/s10052-019-7394-z.
arXiv:1906.05393 [gr-qc].




\bibitem{U9}
T.~S.~Almeida, M.~L.~Pucheu, C.~Romero and J.~B.~Formiga,
``From Brans-Dicke gravity to a geometrical scalar-tensor theory,''
Phys.\ Rev.\ D \textbf{89}, 064047 (2014).
doi:10.1103/PhysRevD.89.064047.

\bibitem{U10}
A.~Kozak and A.~Borowiec,
``Palatini frames in scalar--tensor theories of gravity,''
Eur.\ Phys.\ J.\ C \textbf{79}, 335 (2019).
doi:10.1140/epjc/s10052-019-6836-y.

\bibitem{U11}
K.~Aoki and K.~Shimada,
``Scalar-metric-affine theories: Can we get ghost-free theories from symmetry?''
Phys.\ Rev.\ D \textbf{100}, 044037 (2019).
doi:10.1103/PhysRevD.100.044037.

\bibitem{U12}
L.~L.~Smalley,
``Brans-Dicke-type models with nonmetricity,''
Phys.\ Rev.\ D \textbf{33}, 3590--3593 (1986).
doi:10.1103/PhysRevD.33.3590.


		
	\end{thebibliography}
\end{document}